\documentclass[a4paper,twocolumn,superscriptaddress,11pt,accepted=2018-05-07]{quantumarticle}
\pdfoutput=1
\usepackage[utf8]{inputenc}
\usepackage[english]{babel}
\usepackage[T1]{fontenc}
\usepackage{amsmath}
\usepackage{hyperref}

\usepackage{tikz}
\usepackage{lipsum}

\usepackage{bm,graphicx}
\usepackage{bbm}
\usepackage{dsfont}
\usepackage{amssymb,amsmath,amsfonts}
\usepackage[mathscr]{eucal}
\usepackage{color}

\usepackage[normalem]{ulem}

\definecolor{darkgreen}{rgb}{0,0.6,0}

\newcommand{\id}{\mathds{1}}
\DeclareMathOperator{\tr}{tr}
\DeclareMathOperator{\Tr}{Tr}
\DeclareMathOperator{\diag}{diag}
\newcommand{\ket}[1]{\left|{#1}\right\rangle}
\newcommand{\bra}[1]{\left\langle{#1}\right|}

\newcommand{\rmi}{\ensuremath{\mathrm{i}}}

\newcommand{\gd}{\ensuremath{\mathsf{g}}}
\newcommand{\hd}{\ensuremath{\mathsf{h}}}

\newcommand{\yd}{\ensuremath{\mathsf{y}}}

\newcommand{\swap}{\ensuremath{\mathsf{SWAP}}}
\newcommand{\rhoti}{\ensuremath{\tilde{\rho}}}
\newcommand{\matel}[2]{\left|#1\middle\rangle\middle\langle#2\right|}

\begin{document}
%\noindent
%preliminary 
%draft, \today, please do not distribute\\[5mm]
%
\title
{Distribution of  entanglement and correlations in all finite dimensions}
\author{Christopher Eltschka}
\affiliation{Institut f\"ur Theoretische Physik, Universit\"at Regensburg, 
     D-93040 Regensburg, Germany}
\orcid{}
\author{Jens Siewert}
\affiliation{Departamento de Qu\'{i}mica F\'{i}sica, 
             Universidad del Pa\'{i}s Vasco UPV/EHU, E-48080 Bilbao, Spain}
\affiliation{IKERBASQUE Basque Foundation for Science, E-48013 Bilbao, Spain}
\orcid{}

\begin{abstract}
The physics of a many-particle system is determined 
by the correlations in its quantum state. Therefore, analyzing these 
correlations
is the foremost task of many-body physics. Any `a priori' constraint 
for the properties of the global vs.\ the local states---the so-called
marginals---would help %in this respect 
in order to narrow down the wealth of possible
solutions for a given many-body problem, 
however, little is known about such constraints.
We derive an equality for correlation-related quantities of any
multipartite quantum system composed of finite-dimensional local parties.
This relation defines a necessary condition for the compatibility 
of the marginal properties with those of the joint state.
While the equality holds both for pure and mixed states,
the pure-state version containing only entanglement measures 
represents a fully general
monogamy relation for entanglement.
These findings have interesting implications
in terms of conservation laws for correlations, and also with respect
to topology.
\end{abstract}

%\pacs{03.65.Aa, 03.67.Mn}

\maketitle

%%%%%%%%%%%%%%%%%%%%%%%%%%%%%%%%%%%%%%%%%%%%
%\emph{Introduction}.---
%\section{Introduction}
%%%%%%%%%%%%%%%%%%%%%%%%%%%%%%%%%%%%%%%%%%%%
%
Correlations between different parts of  
many-particle quantum systems are manifestly
different from their classical counterparts.
The hallmark of the latter is that, if the
global state is completely known, so are the
local states, that is, the individual states
of each particle. For quantum states this may
be different: Even if the global state is 
known with certainty, there may be no information
whatsoever regarding the local states such as, for example,
in maximally entangled states of two $d$-dimensional
systems~\cite{Schroedinger1935,Peres}. 

Often, different degrees of departure
from the classical behavior are distinguished, namely
entanglement, steerability, and nonlocality~\cite{Wiseman2007}. 
While all these correlation types are based on entanglement,
nonlocal correlations are regarded as the strongest. This
is because all nonlocal states are entangled, but only
those for which there are no local models may violate
a Bell inequality~\cite{Werner1989,Brunner2014}. 
Already such basic considerations
reveal that there must exist mathematical tools to characterize
quantitative aspects of correlations in a multipartite quantum state.
Entanglement measures~\cite{Horodecki2009,ES2014} constitute one example 
of such tools, but there
are many others~\cite{Streltsov2015,Adesso2016} 
and to date it is not clear how to exhaustively
characterize quantum correlations, not even for finite-dimensional 
systems.

Such a characterization can be achieved via mathematical constraints
for different correlation functions and/or quantifiers. A Bell inequality
is precisely one such constraint: It relates the existence of a
local model to the maximum absolute value of a certain 
linear combination of correlation functions. Another example are
so-called monogamy inequalities for entanglement~\cite{CKW2000,KW2004,Osborne2006,Bai2014,Adesso2014}
or nonlocality~\cite{Toner2009,Seevinck2010} which establish constraints 
between the correlations that may coexist in different subsystems 
of a multipartite quantum system.

It is remarkable that there exist {\em exact} relations, 
that is, {\em equalities} between certain entanglement measures
in qubit systems. Coffman  et al.~\cite{CKW2000}
discovered the now famous relation for pure states $\psi_3$ 
of three qubits (denoted by $A$, $B$, and $C$)
\begin{align}
   \tau_3(\psi_3)\ =\ \tau_{A}-C_{AB}^2\ -\ C_{AC}^2\ \ \ ,
\label{eq:CKW}
\end{align}
where $\tau_A$ denotes the linear entropy of the reduced
state $\rho_A=\Tr_{BC}\ket{\psi_3}\!\bra{\psi_3}$, $C_{AB}$ is the
concurrence of the two-qubit reduced state 
$\rho_{AB}=\Tr_C\ket{\psi_3}\!\bra{\psi_3}$ (analogously for $C_{AC}$),
and $\tau_3(\psi_3)$ is the absolute value of
a polynomial local SL (LSL) invariant that depends on the
state coefficients  of $\psi_3$. It turned out only recently~\cite{ES2015} that
similar relations 
%(in fact, very likely there is a hierarchy of them) 
exists for any $N$-qubit system, the simplest being
\begin{align}
   2|H(\psi_N)|^2 = \sum_{k=1}^N (-1)^{k+1}\sum_{j_1< \ldots <  j_k} 
                                 \tau_{A_{j_1}\ldots A_{j_k}}\ ,
\label{eq:monog2}
\end{align}
where now $\tau_{A_{j_1}\ldots A_{j_k}}$ is the linear entropy
of the reduced state of qubits $A_{j_1}, \ldots, A_{j_k}$
and $H(\psi_N)=\bra{\psi_N^*} \sigma_2^{\otimes N}\ket{\psi_N}$ 
is another polynomial LSL invariant~\cite{Wong2001,ES2014} 
(here $\ket{\psi_N^*}$ is
the complex conjugate of $\ket{\psi_N}$ and $\sigma_2$ is the
second Pauli matrix). Strikingly, all of the quantities in
Eqs.~\eqref{eq:CKW},~\eqref{eq:monog2} are 
measures for entanglement, either of the individual party (the linear
entropies and the concurrences) or in the global state (the polynomial
LSL invariants). These relations impose rigid constraints between
the properties of the global and the reduced states, and are therefore
intimately connected with the 
{\em quantum marginal problem}~\cite{Klyachko2006}. However, since
their validity appears to be restricted to qubit systems, it is not
clear whether they bear any relevance to a general physical setting. 

This is the motivation for the present work: We show that the analogue
of Eq.~\eqref{eq:monog2} holds for {\em any} multipartite system
of finite-dimensional constituents. 
In passing, by this relation a local unitary invariant of the
pure state is introduced, which we term {\em distributed concurrence},
because it describes the 
distribution of bipartite entanglement in
the pure state and is a natural generalization both of 
the bipartite concurrence and of $|H(\psi_N)|$ in Eq.~\eqref{eq:monog2}. 
%and, as we prove, 
%an entanglement monotone.

%%%%%%%%%%%%%%%%%%%%%%%%%%%%%%%%%%%%%%%%%%%%
\section{Universal state inversion}
%\emph{Universal state inversion}.---
%%%%%%%%%%%%%%%%%%%%%%%%%%%%%%%%%%%%%%%%%%%%
%
\subsection{Definition}
%%%%%%%%%%%%%%%%%%%%%%%%%%%%%%%%%%%%%%%%%%%%

The key ingredient in our investigation is the operator 
obtained by applying the so-called 
{\em universal state inversion}~\cite{Horodecki1999,Rungta2001},
see also Refs.~\cite{Hall2005,Hall2006,Breuer2006,Lewenstein2016}.
Consider a $d$-dimensional Hilbert space $\mathcal{H}$ and 
$\mathcal{B}(\mathcal{H})$,
the set of bounded positive semidefinite 
operators on $\mathcal{H}$. Horodecki 
et al.~\cite{Horodecki1999} defined the operator
\begin{align}
          S_d(\mathcal{O})\ =\ \Tr(\mathcal{O})\ \id\ -\ \mathcal{O}
          \ \ \ ,\ \ 
          \mathcal{O}\in\mathcal{B}(\mathcal{H})
\label{eq:inv}
\end{align}
which was used by Rungta et al.~\cite{Rungta2001} to write
the concurrence of a state $\psi \in \mathcal{H}_A\otimes \mathcal{H}_B$ as
\begin{align}
          C(\psi)\ =\ & \sqrt{
                \bra{\psi} S_{d_A}\otimes S_{d_B}(\ket{\psi}\!\bra{\psi})
                                                  \ket{\psi}
                            }
\label{eq:inv2conc1}
\\
                   =\ & \sqrt{2-\Tr_A \rho_A^2-\Tr_B\rho_B^2 
                              }\ \ ,
\label{eq:inv2conc2}
\end{align}
(where $d_j=\mathrm{dim}\ \mathcal{H}_j$, $j=A,B$; 
$\rho_A=\Tr_B\ket{\psi}\!\bra{\psi}$, 
$\rho_B=\Tr_A\ket{\psi}\!\bra{\psi}$ 
are the reduced states for the parties).
They termed the operator $S_d$ the {\em universal state inverter}.
We will show below that, 
for non-Hermitian operators $\mathcal{O}$, the appropriate definition of the
single-system state inverter is 
$\tilde{\mathcal{O}}=
\Tr \left(\mathcal{O}^{\dagger}\right)\id-\mathcal{O}^{\dagger}$ 
(see Section \ref{sec:generators}).

%Here, our objective is to  emphasize the action of 
%the state inverter on local systems.
It is straightforward to generalize universal state inversion toward
density matrices $\rho$ of $N$-partite systems, i.e., 
$\rho \in \mathcal{B}(\mathcal{H}_1\otimes\ldots\otimes\mathcal{H}_N)$,
$X\in \{1,\ldots,N\}$
\begin{align}
  \tilde \rho & =   
                    \mathcal S_{d_1}\mathcal S_{d_2}
                    \ldots\mathcal S_{d_n}(\rho)
\label{eq:rhotilde}\\
              & = \Big[ \prod_{X=1}^N \big( \Tr_X(\cdot)\otimes\id_X 
                              - \id 
                                     \big)
                  \Big]\ \rho
                            \ \ ,
\label{eq:rhotildeprod}
\end{align}
where $\id_X$ denotes the identity on a single subsystem $X$, 
as opposed to $\id$, the identity on the full system.
We use the tilde notation for the inverted state $\tilde{\rho}$
in order to make reference to Wootters' notation in Ref.~\cite{Wootters1998},
Eq.~(5), because that definition for two qubits is naturally generalized 
by Eq.~\eqref{eq:rhotildeprod}, as we will see below.
%the two-qubit case considered by Wootters. 
Note that if the dimension
of any party is larger than 2, the trace of $\rhoti$ is larger than 1.
Nonetheless we will call $\rhoti$ the `tilde state' if no ambiguity
is possible. 

\begin{figure}[htb]
  \centering
  \includegraphics[width=.97\linewidth]{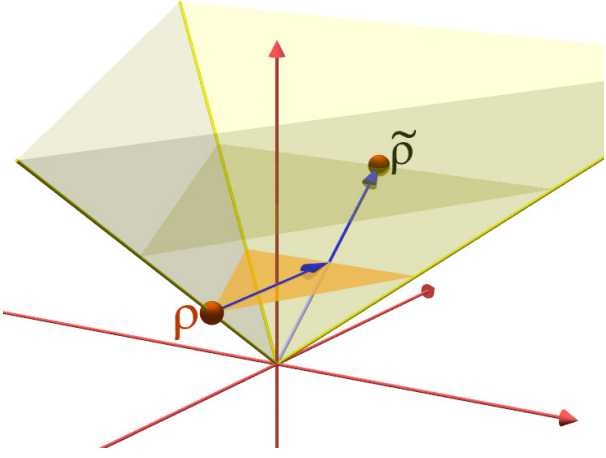}
  \caption{Illustration of the action of the state inverter 
           on a pure state $\rho$ in dimension $d=3$. 
           The orange triangle is a section through the
           qutrit Bloch body containing the pure state $\rho$ 
           at one corner (orange sphere). 
           The other two corners of the triangle are
           pure states of which $\rhoti$ is a linear combination. The vertical
           axis denotes the multiples of the identity matrix determining 
           the trace. All density matrices have trace $1$, thus the Bloch
           body is located at a constant height. The axis crosses the Bloch body
           in the completely mixed state.
%          , which in this section is in the center
%          of the triangle.
           The state inverter maps the pure state to the state lying furthest on
           the opposite site of the completely mixed state, as shown by the blue
           arrow crossing the vertical axis. This state is an equal mixture of
           the other two pure states in the triangle. At the same time,
           it scales the vector by the factor $(d-1)=2$, as indicated by the
           second blue arrow. This scaling maps the Bloch body to a copy at
           height $(d-1)$, shown as semitransparent yellow triangle.
           The resulting operator $\rhoti$ is again marked by a small orange
           sphere.
    }
  \label{fig:invert-action}
\end{figure}

The operator $\rhoti$ appears quite frequently in algebraic considerations
regarding properties of quantum states, in particular, entanglement.
Yet it is not obvious to attribute to it a physical
significance beyond the interpretation of Eq.~\eqref{eq:inv}
%for the single-system inverter 
given in Ref.~\cite{Rungta2001}:
It takes a pure state to the maximally mixed state in the subspace
orthogonal to the pure state, multiplied by the dimension of
that subspace. The latter statement can be re-expressed
in the following way: Geometrically, the single-system inverter maps 
a pure state of a $d$-dimensional system to the barycenter of the 
hypersurface opposite to it, that is, the convex hull of those pure
states which, together with the original state, form an orthonormal
basis. The result is then scaled with a prefactor $(d-1)$, 
cf.~Fig.~\ref{fig:invert-action}.

%It is easy to see that 
This geometrical interpretation can be extended
to the multipartite case. To this end, consider a state of the 
computational basis $\ket{j_1\ldots j_N}\in \mathcal{H}_1\otimes\cdots\otimes
\mathcal{H}_N$. Then we have
\begin{align}
      \widetilde{\ket{j_1\ldots j_N}\!\bra{j_1\ldots j_N}} = 
                  \bigotimes_{k=1}^N 
                                          \big(
                                    \id_{k}- \ket{j_k}\!\bra{j_k}
                                          \big)
 \ ,
\end{align}
which corresponds to the barycenter of the hypersurface formed by
all those states of the multipartite 
computational basis whose
$k$-th tensor factor is different from $\ket{j_k}$. 
%This state
%is multiplied by the total dimension of that 
%hypersurface, $\prod_{k=1}^N (d_k-1)$.
This state is multiplied by a factor $\prod_{k=1}^N (d_k-1)$ 
that keeps track of the total dimension of that 
hypersurface. 
%%%%%%%%%%%%%%%%%%%%%%%%%%%%%%%%%%%%%%%%%%%%%%%%%%%%%%%%%%%%%%%%%%%%%
%% FIGURE 1
%%%%%%%%%%%%%%%%%%%%%%%%%%%%%%%%%%%%%%%%%%%%%%%%%%%%%%%%%%%%%%%%%%%%%
%%%%%%%%%%%%%%%%%%%%%%%%%%%%%%%%%%%%%%%%%%%%%%%%%%%%%%%%%%%%%%%%%%%%%
%

\subsection{Representation in terms of reduced states}
%%%%%%%%%%%%%%%%%%%%%%%%%%%%%%%%%%%%%%%%%%%%%%%%%%%%%%%%%%%

There are several noteworthy ways to explicitly write the state $\tilde{\rho}$.
The first and obvious possibility is %connect the definitions~\eqref{eq:inv}
to expand the operator product in Eq.~\eqref{eq:rhotildeprod} which leads
to an alternating sum over all bipartitions $A$~\cite{Hall2005} 
\begin{align}
     \tilde{\rho}\ =\ \sum_{A}\ (-1)^{|A|}\ \left(\Tr_{\bar{A}}\rho
                                            \right)\otimes 
                          \id_{\bar{A}}
\ \ .
\label{eq:rhotilde-reduced}
\end{align}
Here $A$ denotes a subset of $\{1,\ldots,N\}$, $|A|$ is the number of
elements in $A$, $\bar{A}$ is the complement of the subset $A$,
and $\rho_A=\Tr_{\bar{A}}\rho$ is the reduced density matrix of the parties 
in the subset $A$.

\subsection{Relation to the Bloch representation of the original state}
%%%%%%%%%%%%%%%%%%%%%%%%%%%%%%%%%%%%%%%%%%%%%%%%%%%%%%%%%%%%%%%%%%%%%%%%%%

A second important way to write $\tilde{\rho}$ is its Bloch representation.
In order to obtain it we first consider the Bloch representation of 
$\rho$ itself. An $N$-qudit state $\rho$ can be written as
\begin{align}
    \rho\ =\ & \frac{1}{d_1\cdots d_N}\sum_{j_1 \ldots j_N}
                      r_{j_1\ldots j_N}\ \gd_{j_1}\otimes\cdots\otimes\gd_{j_N}
\label{eq:bloch-qudit}
\\        =\ & \frac{1}{d_{\mathrm{tot}}}\left( \id
                           + P_1 + P_2 + \ldots + P_{N}
                                         \right)
\label{eq:bloch-quditP}
\ \ .
\end{align}
Here, $\{\gd_{j_k}\}$ are sets of $d_k^2-1$ %\tcr{\sout{Hermitian}} 
traceless generators
of SU($d_k$), $\gd_{j_k}=\id_{k}$ for $j_k=0$, $r_{j_1\ldots j_N}\in 
\mathbb{C}$. Moreover, $d_{\mathrm{tot}}=d_1\cdots d_N$ is the total dimension
of the Hilbert space. We use the symbol $P_q$ for the sum of all
those terms in Eq.~\eqref{eq:bloch-qudit}
that contain the same number $q$ of notrivial generators.

The Bloch representation of $\tilde{\rho}$ is particularly nice if all
subsystem dimensions are equal, i.e., $d_1=d_2=\ldots =d_N\equiv d$
and $d_{\mathrm{tot}}=d^N$. In order
to avoid cumbersome notation we restrict ourselves to this case. Then it
is easy to see by directly applying the product form 
of the state inversion operator Eq.~\eqref{eq:rhotildeprod}
to each term in the Bloch representation
Eq.~\eqref{eq:bloch-quditP} separately that
\begin{align}
    \rhoti\ = &\ 
        \frac{1}{d^N}\Big[\ (d-1)^N\id\
                           -\ (d-1)^{N-1}\ P_1\  + \Big.
\nonumber\\  &  \ \  \  +\ (d-1)^{N-2}P_2\ -\ +\ \ldots\  + 
\nonumber\\  &  \ \  \ +
              \Big.   (-1)^{N-1}(d-1) P_{N-1} + (-1)^{N} P_N
                                         \Big]
\label{eq:rhotildeP}
 \ .
\end{align}
It is obvious from Eq.~\eqref{eq:rhotildeP} that for $d>2$ the universal
state inverter is not a trace-preserving map. Further, it shows that
only for $d=2$ the tilde state of a pure state $\Pi_{N\mathrm{qubit}}$
is pure %and normalized 
as well, because in that case we have for the purity of the tilde state
\begin{align*}
      \big(\Tr \tilde{\Pi}_{N\mathrm{qubit}}\big)^2\ =\  &
      \Tr \tilde{\Pi}_{N\mathrm{qubit}}^2
\\
                                              \ =\  &
      \frac{1}{2^N}
                 \sum_{j_1\ldots j_N} \big| r_{j_1\ldots j_N} \big|^2
\\                                            \ =\  & 1\ \ .
\end{align*}

\subsection{Explicit generator representation}
%%%%%%%%%%%%%%%%%%%%%%%%%%%%%%%%%%%%%%%%%%%%%%%%%%%%%%%%%%%%%%%%%%%%%%%%%%
\label{sec:generators}

Finally, we want to find a representation of $\rhoti$ in terms of
generators of the unitary group. To this end, we use the
generalized Gell-Mann matrices~\cite{Georgi1982} in $d$ dimensions
\begin{subequations}
\begin{align}
  \label{eq:genx}
  \mathsf{x}_{kl}^{(d)} &=
  \sqrt{\frac{d}{2}}(\matel{k}{l}+\matel{l}{k}) \\
%  & (0\leq k<l<d)\\
  \label{eq:geny}
  \mathsf{y}_{kl}^{(d)} &=
  \sqrt{\frac{d}{2}}(-\mathrm i\matel{k}{l}+\mathrm i\matel{l}{k}) \\
%  & (0\leq k<l<d)\\
  \label{eq:genz}
  \mathsf{z}_l^{(d)} &= \sqrt{\frac{d}{l(l+1)}}
  \left(-l\matel{l}{l} + \sum_{j=0}^{l-1}\matel{j}{j}\right) 
%& (0<l<d)
\end{align}
\end{subequations}
where $0\leqq k < l < d$.
The normalization is chosen such that
\begin{equation}
  \label{eq:xyznormalization}
  \begin{aligned}
    \Tr(\mathsf{x}_{kl}\mathsf{x}_{mn}) &=
    \Tr(\mathsf{y}_{kl}\mathsf{y}_{mn}) = d\ \delta_{km}\delta_{ln}\ \ ,\\
    \Tr(\mathsf{z}_l\mathsf{z}_n) &= d\ \delta_{ln}\ \ .
%\\
%    \tr(\mathsf{x}_{kl}\mathsf{y}_{mn}) &=
%    \tr(\mathsf{x}_{kl}\mathsf{z}_{n}) =
%    \tr(\mathsf{y}_{kl}\mathsf{z}_{n}) = 0
  \end{aligned}
\end{equation}
%
%for any $0\leq k<l<d$ and $0\leq m<n<d$. 
Note that, up to a factor, these are 
the usual $d$-dimensional Gell-Mann matrices.

Just as above in Eq.~\eqref{eq:bloch-qudit} it is useful 
to include the identity matrix, as this
gives a complete basis for the Hermitian matrices. %% if using real
%coefficients, or for all matrices if using complex matrices. Also,
Sometimes we will use a uniform labeling with only one
index and therefore introduce also the alternative form
\begin{subequations}
\begin{align}
  \label{eq:generators}
  \mathsf{h}_0^{(d)} &\ = \ \id\\
  \mathsf{h}_{l^2+2k}^{(d)} &\ = \ \mathsf{x}_{kl}\\
  \mathsf{h}_{l^2+2k+1}^{(d)} &\ =\  \mathsf{y}_{kl}\\
  \mathsf{h}_{l^2+2l}^{(d)} &\ =\  \mathsf{z}_l
\end{align}
\end{subequations}
where the index corresponds to the usual numbering of the Gell-Mann
matrices. Relations \eqref{eq:xyznormalization} then simplify to
\begin{equation}
  \label{eq:hnormalization}
  \Tr(\mathsf{h}_k\mathsf{h}_l)\  =\  d\ \delta_{kl} \ \ .
\end{equation}
For the following considerations we need two formulae
for the single-system density matrix $\rho \in \mathcal{B}(\mathcal{H})$
with $\dim\mathcal{H}=d$
(a proof is given in Appendix~A)
\begin{align}
  \label{eq:tracefromgenerators}
  (\Tr \rho)\id &\ =\  \frac{1}{d}\sum_{k=0}^{d^2-1} \mathsf{h}_k\rho\mathsf{h}_k\\
  \label{eq:transposefromgenerators}
  \rho^T &\ =\  \frac{1}{d}\sum_{k=0}^{d^2-1} \mathsf{h}_k^T\rho\mathsf{h}_k
\ \ .
\end{align}
%
%as well as $\rho = \rho^\dagger = (\rho^T)^*$ and w$\tr \rho = (\tr \rho)^*$. 
%
By applying these relations to Eq.~\eqref{eq:inv} we find
for the tilde state of a single system
\begin{align}
  \label{eq:inverterop}
  \tilde \rho &\ = \ (\Tr \rho)\id - \rho     \nonumber\\
  &\ =\ \left[(\Tr \rho)\id - \rho^T\right]^*     \nonumber\\
  &\ =\ \frac{1}{d}\sum_{k=0}^{d^2-1} 
        \left[\mathsf{h}_k\rho \mathsf{h}_k - 
                     \mathsf{h}_k^T\rho\mathsf{h}_k\right]^* \ \ .
\end{align}
To proceed from here, we note that the generators $\mathsf{y}_{kl}$
change their sign on transposition, while all other generators remain
unchanged under that operation. Therefore the sum simplifies to
\begin{align}
  \tilde \rho &\ =\
  \frac{2}{d}\sum_{k=0}^{d-2}\sum_{\ l=k+1}^{d-1}
                         (\mathsf{y}_{kl}\, \rho\, \mathsf{y}_{kl})^* 
\nonumber\\
              &\ =\ 
  \frac{2}{d}\sum_{k=0}^{d-2}\sum_{\ l=k+1}^{d-1}
                          \mathsf{y}_{kl}\, \rho^*\, \mathsf{y}_{kl}
  \label{eq:rhotildegenerators}
\ \ .
\end{align}
%
%where the last equality is because the complex conjugation of the
%$\mathsf{y}$ terms only cause sign changes that cancel out.

It is straightforward to extend the result~\eqref{eq:rhotildegenerators}
to states of multipartite systems 
$\rho\in \mathcal{B}(\mathcal{H}_1\otimes\cdots\otimes\mathcal{H}_N)$.
For this purpose, we introduce the simplified notation
\begin{align*}
% \label{eq:multiindex}
  k &= (k_1,\ldots,k_N)   \\
%  d_* &= d_1d_2 \cdots d_N   \\
  \mathsf{y}_{kl} &=
  \mathsf{y}_{k_1l_1}\otimes\mathsf{y}_{k_2l_2}\otimes\cdots\otimes\mathsf{y}_{k_Nl_N}   \nonumber\\
  \sum_{k<l}^{d-1} &\equiv \sum_{k_1=0}^{d_1-2}\sum_{l_1=k_1+1}^{d_1-1} \cdots
  \sum_{k_N=0}^{d_N-2}\sum_{l_N=k_N+1}^{d_N-1}   
\end{align*}
and can then write
\begin{align}
    \rhoti\ =\ \frac{2^N}{d_{\mathrm{tot}}}\ \sum_{k<l}^{d-1}
                             \yd_{kl}\ \rho^*\ \yd_{kl}
\ \ .
\label{eq:rhotildegen}
\end{align}
This is a remarkable result, not only because of its simplicity.
It makes explicit that universal state inversion is an
antilinear operation. In fact, $\rhoti$ in Eq.~\eqref{eq:rhotildegen}
is in all respects the natural
generalization of Wootters' tilde state.  We note also that for pure states
$\rho_{\psi}=\matel{\psi}{\psi}\equiv \Pi_{\psi}$ 
the tilde state $\tilde \Pi_{\psi}$ in general is mixed,
only for $d_k=2$ the generator sum contains a single term, 
and $\tilde \Pi_{\psi}$ becomes pure as well.

Finally, it is immediately clear from Eq.~\eqref{eq:rhotildegen} that the
tilde state $\rhoti$ is positive: With $\rho$, also $\rho^*$ is
positive and therefore $\rho^*=\sqrt{\rho^*}\cdot\sqrt{\rho^*}$, so
that 
\begin{align}
    \rhoti\ & =\ \frac{2^N}{d_{\mathrm{tot}}}\ \sum_{k<l}^{d-1}
                             \yd_{kl}\ \sqrt{\rho^*}\sqrt{\rho^*}\ \yd_{kl}
\nonumber\\
            & =\ 
     \frac{2^N}{d_{\mathrm{tot}}}\ \sum_{k<l}^{d-1}
                 \left(\sqrt{\rho^*}\ \yd_{kl}\right)^{\dagger}\ 
                         \sqrt{\rho^*}\ \yd_{kl}
\ \ .
\label{eq:positive}
\end{align}
That is, $\rhoti$ is a sum of positive operators and therefore also positive.
An important consequence of Eq.~\eqref{eq:positive} is 
\begin{align}
    \Tr\left(\rho\rhoti\right)\ =\ 
        \Tr\left[\left(\sqrt{\rhoti}\sqrt{\rho}
                 \right)^{\dagger}
                 \left(\sqrt{\rhoti}\sqrt{\rho}
                 \right)
           \right]\ \geqq\ 0\ \ .
\label{eq:TrRpos}
\end{align}

%%%%%%%%%%%%%%%%%%%%%%%%%%%%%%%%%%%%%%%%%%%%%%%%%%%%%%%%%%%%%%
\section{Distributed concurrence}
%%%%%%%%%%%%%%%%%%%%%%%%%%%%%%%%%%%%%%%%%%%%%%%%%%%%%%%%%%%%%%
\subsection{New concurrence-type invariant}
%%%%%%%%%%%%%%%%%%%%%%%%%%%%%%%%%%%%%%%%%%%%%%%%%%%%%%%%%%%%%%

By using the notation $\matel{\psi}{\psi}\equiv \Pi_{\psi}$
we can express the concurrence Eq.~\eqref{eq:inv2conc2} for the
pure state of a bipartite system as 
%$\psi\in \mathcal{H}_1\otimes\mathcal{H}_2$,
%
%\begin{align}
$   C(\psi)\ =\ \sqrt{ \Tr \big(\Pi_{\psi} \tilde \Pi_{\psi} \big) } $.
%\end{align}
%
It is straightforward to extend this definition to pure states
of multipartite systems, $\psi\in \mathcal{H}_1\otimes\cdots\mathcal{H}_N$.
We try
\begin{align*}
   C_D(\psi)^2\ & =\ \Tr \big(\Pi_{\psi} \tilde \Pi_{\psi} \big)  
\\
              & =\ \frac{2^N}{d_{\mathrm{tot}}}\ \sum_{k<l}^{d-1}
                      \Tr \big(  \Pi_{\psi} \yd_{kl}\ \Pi_{\psi}^*\ \yd_{kl}
                          \big)
\\
              & =\ (-1)^N\frac{2^N}{d_{\mathrm{tot}}}\ \sum_{k<l}^{d-1}
                      \left|\bra{\psi^*}\yd_{kl}\ket{\psi}\right|^2 
\ \ .
\end{align*}
Here we see that for odd $N$, the quantity $C_D(\psi)$ vanishes
identically, for any local dimension, in analogy with the behavior
of the $N$-qubit LSL invariant 
$H(\phi_N)=\bra{\phi^*_N}\sigma_2^{\otimes N} \ket{\phi_N}$ 
(where $\phi_N\in \mathbb{C}_2^{\otimes N}$). 
This is because $\bra{\phi^*_N}\yd_{kl} \ket{\phi_N}
                =\bra{\phi_N}\yd_{kl}^* \ket{\phi^*_N}^*
                =\bra{\phi^*_N}(-\yd_{kl}) \ket{\phi_N}\equiv 0$
for an odd number $N$ of factors in the tensor product $\yd_{kl}$.

We can define now the {\em distributed concurrence}
\begin{align}
   C_D(\psi)\ & =\ \sqrt{\Tr \big(\Pi_{\psi} \tilde \Pi_{\psi} \big) }
\label{eq:cdquadrat}
\\
              & =\ \sqrt{\frac{2^N}{d_{\mathrm{tot}}}\ \sum_{k<l}^{d-1}
                      \left|\bra{\psi^*}\yd_{kl}\ket{\psi}\right|^2 }
\label{eq:CDy}
\ \ .
\end{align}
Since the universal state inverter~Eq.~\eqref{eq:rhotilde}
 commutes with local unitary
operations~\cite{Rungta2001}, the distributed concurrence
is a local unitary invariant.
We will show below that $C_D(\psi)$ in certain cases is also an
entanglement monotone.

Again, the quantity $C_D(\psi)$ generalizes the known concurrence-type
quantities in the most natural way: For $N$-qubit states $\phi_N$
the LSL invariant $|H(\phi_N)|$ is obtained, 
in particular for $N=2$ we get
Wootters' concurrence $|\bra{\phi_2^*}\sigma_2\otimes\sigma_2\ket{\phi_2}|$.
On the other hand, for two parties in dimensions  $d_1=d_2>2$ 
we get the standard $d\times d$ concurrence, Eq.~\eqref{eq:inv2conc2}
(cf.\ also Ref.~\cite{Albeverio2001,Akhtarshenas2005,Li2008}).

Concluding this section, we note that the operator 
$\Pi_{\psi}\tilde\Pi_{\psi}\Pi_{\psi}$ is proportional to $\Pi_{\psi}$
because of the projector property of $\Pi_{\psi}$. Then, by virtue
of Eq.~\eqref{eq:cdquadrat}, we have
\begin{align}
   \Pi_{\psi}\tilde\Pi_{\psi}\Pi_{\psi}\ =\ C_D(\psi)^2\ \Pi_{\psi}
\ \ . 
\end{align}
%
%%%%%%%%%%%%%%%%%%%%%%%%%%%%%%%%%%%%%%%%%%%%%%%%%%%%%%%%%%%%%%
\subsection{Distributed concurrence is an entanglement monotone
            for local dimensions smaller than four}
\label{sec:monotonie}
%%%%%%%%%%%%%%%%%%%%%%%%%%%%%%%%%%%%%%%%%%%%%%%%%%%%%%%%%%%%%%

In order to investigate the monotone property 
of the distributed concurrence
we need to study whether or not, for pure states, $C_D(\psi)$ can 
increase under the action of separable Kraus operators 
$O_j=A_1^{(j)}\otimes\ldots\otimes A_N^{(j)}$ 
(where $\sum_j O_j^{\dagger}O_j=\id$), which represent
stochastic local operations and classical communication
(SLOCC)~\cite{Vidal1999,DVC2000}. 
If $C_D$ is an entanglement monotone for pure states,
it can be extended to the mixed states, as usual,
through the convex roof~\cite{Uhlmann2010}. 

As the distributed concurrence
acts symmetrically on the parties it suffices to consider
actions on the first party only, 
$\mathcal{A}_j=A_j\otimes\id_2\otimes\ldots\otimes \id_N$. 
We may assume that the number of parties $N$ is even and that 
the reduced density matrix of the first party has rank $r_1 > 1$
because otherwise the monotone property is trivially valid.
Further, since all generalized measurements may be composed from
two-outcome measurements, we will consider Kraus operators
with $A_1^{\dagger}A_1+A_2^{\dagger}A_2=\id$. Then, for
$\psi\in \mathcal{H}_1\otimes\cdots\otimes\mathcal{H}_N$ 
we need to show
\begin{align}
   C_D(\psi)\ & \geqq\ p_1\ C_D\left(\frac{\mathcal{A}_1\psi}
                                        {\sqrt{p_1}}
                             \right)\ +\
   p_2\ C_D\left(\frac{\mathcal{A}_2\psi}
                                        {\sqrt{p_2}}
                             \right)
\nonumber\\
              & \geqq\ 
   C_D\left(\mathcal{A}_1\psi \right)\ +\
   C_D\left(\mathcal{A}_2\psi
                             \right)
\label{eq:mon1}
\end{align}
with $p_j=\Tr\left(\mathcal{A}_j\Pi_{\psi}\mathcal{A}_j^{\dagger}\right)$. 
In the second line we have used the property $C_D(\alpha\psi)=
|\alpha|^2 C_D(\psi)$. 

We now rewrite Eq~\eqref{eq:mon1} by using the decomposition
$\Pi_{\psi}=\sum_{jk}\ket{e_jf_j}\!\bra{e_kf_k}$, where 
$\{\ket{e_j}\}$ are orthonormal vectors that span the Hilbert space of the
first party and are chosen such that they diagonalize the operator
$\mathcal{A}_1^{\dagger}\mathcal{A}_1$. The vectors
$\{\ket{f_j}\}$ span the Hilbert space of the last $(N-1)$ parties in $\psi$
and are not necessarily orthogonal or normalized, but they obey
$\sum_j\bra{f_j}f_j\rangle=1$.
We obtain
\begin{align*}
     \tr \big(\Pi_{\psi}\tilde{\Pi}_{\psi}\big)
     \ =\ & 2\sum_{jk}  F_{kjjk}
\\
     \tr \big(A_1\Pi_{\psi}A_1^{\dagger}
         \big)
         \widetilde{
         \big(A_1{\Pi}_{\psi}A_1^{\dagger}
         \big)
                   }
               \ =\ & 2\sum_{kl} F_{kjjk} D_j D_k
\end{align*}
where we have used the definitions
\begin{align*}
 D_k \delta_{jk}\equiv\ & \bra{e_k}\mathcal{A}_1^{\dagger}\mathcal{A}_1\ket{e_j}
\ \ , \ 0\leqq D_k\leqq 1\ ,
\\
 F_{klmj}\ \equiv\ & \bra{f_k}\left(\widetilde{\ket{f_l}\!\bra{ f_m }}
                              \right)
                     \ket{f_j}
\ \ .
\end{align*}
Note that $F_{klmj}=F_{lkjm}=-F_{lkmj}$, which follows with the explicit
inversion formula~\eqref{eq:rhotildegen}. 
%%We also have eliminated $\mathcal{A}_2$ by
%%using $\mathcal{A}_2^{\dagger}\mathcal{A}_2=\id-\mathcal{A}_1^{\dagger}
%%\mathcal{A}_1$. 
Finally, $p_1=\Tr\left(\mathcal{A}_1\Pi_{\psi}
\mathcal{A}_1^{\dagger}\right)=\sum_j D_j \bra{f_j}f_j\rangle$ and 
         $p_2=\Tr\left(\mathcal{A}_2\Pi_{\psi}
\mathcal{A}_2^{\dagger}\right)=\sum_j (1-D_j)\bra{f_j}f_j\rangle$. 

With all this, we can transform Eq.~\eqref{eq:mon1} into the
equivalent inequality
\begin{align}
 \sqrt{\sum_{jk}  F_{jkkj}} & \ \geqq\ 
          \sqrt{\sum_{jk}  F_{jkkj}D_jD_k}\ +\
\nonumber\\
   & \  +\ \sqrt{\sum_{jk}  F_{jkkj}(1-D_j)(1-D_k)}
\ .
\label{eq:mon2}
\end{align}
Now it is straightforward to prove 
the monotone property for the well-known cases $d=2$, $N$ arbitrary, and 
$N=2$, $d$ arbitrary, as we show in Appendix~\ref{app:monotones}. 
For the general case, we continue transforming Eq.~\eqref{eq:mon2}
by subtracting the second term on the right, squaring and simplifying
the result, to obtain
\begin{align}
      \left( \sum_{jk} w_{jk}D_j \right)^2\ \geqq\ \sum_{jk} w_{jk} D_j D_k
\ \ ,
\label{eq:mon3}
\end{align}
with the weights $w_{jk}\equiv F_{jkkj}/
                          \left(\sum_{lm}F_{lmml}\right)$.
Note that, if the denominator here were zero, the monotone property would
be trivially satisfied.
It is possible to construct examples of $\{D_k\}$ and
$\{ F_{lmml} \}$ such that inequality~\eqref{eq:mon3} is violated
for $\dim{\mathcal{H}_1}=4$ (see Appendix~\ref{app:counterex}).
Therefore, $C_D$ is {\em not} an entanglement monotone if a party has
local dimension $d_j\geqq 4$ and the party number $N\geqq 4$. 
Similarly, such sets can be found to show
that $C_D^2$ is not a monotone for $d_j\geqq 3$, $N\geqq 4$.

Remarkably, however, Eq.~\eqref{eq:mon3} does hold for 
$\dim{\mathcal{H}_1}=3$,
that is, $C_D$ is an entanglement monotone for any number
of parties if all local dimensions
are not larger than three.
This behavior resembles that of LSL invariants for which it is known
that the monotone property depends on the homogeneity 
degree~\cite{Eltschka2012} 
and, in fact, also on the local dimension.

For the proof of Eq.~\eqref{eq:mon3} in the case $\dim{\mathcal{H}_1}=3$
we collect only the $D_j^2$ terms on the left and multiply both
sides by 4. Moreover, we use $2(w_{01}+w_{02}+w_{12})=1$ and obtain
\begin{widetext}
\begin{align}
 (  1- 2w_{12})^2  D_0^2+(1-2w_{02})^2D_1^2+
    (1-2w_{01})^2 D_2^2  \geqq
     &   D_0 \sqrt{2(2w_{01})-(1-2w_{12})(1-2w_{02})}^2 D_1\ +
\nonumber\\
     & \!\! + D_0 \sqrt{2(2w_{02})-(1-2w_{12})(1-2w_{01})}^2 D_2\ +
\nonumber\\
     & \!\! + D_1 \sqrt{2(2w_{12})-(1-2w_{01})(1-2w_{02})}^2 D_2\ +
\nonumber\\
     & \!\! +\ D_1 \sqrt{2(2w_{01})-(1-2w_{12})(1-2w_{02})}^2 D_0\ +
\nonumber\\
     & \!\! + D_2 \sqrt{2(2w_{12})-(1-2w_{02})(1-2w_{01})}^2 D_1\ +
\nonumber\\
     & \!\! + D_2 \sqrt{2(2w_{02})-(1-2w_{12})(1-2w_{01})}^2 D_0 \ .
\label{eq:mon4}
\end{align}
\end{widetext}
The peculiar way of writing Eq.~\eqref{eq:mon4} indicates how one can see
that this is precisely a Cauchy-Schwarz inequality, and therefore is valid.
This concludes the proof that Eq.~\eqref{eq:mon3} holds for 
$\dim\mathcal{H}_1=3$.

To summarize this section (see also Fig.~2), 
we have found that both $C_D$ and $C_D^2$
are entanglement monotones only in the well-known cases and, trivially,
for odd number of parties $N$.
%
%%%%%%%%%%%%%%%%%%%%%%%%%%%%%%%%%%%%%%%%%%%%%%%%%%%%%%%%%%%%%%%%%%%%%
%% FIGURE 2
%%%%%%%%%%%%%%%%%%%%%%%%%%%%%%%%%%%%%%%%%%%%%%%%%%%%%%%%%%%%%%%%%%%%%
\begin{figure}[h]
  \centering
  \includegraphics[width=.99\linewidth]{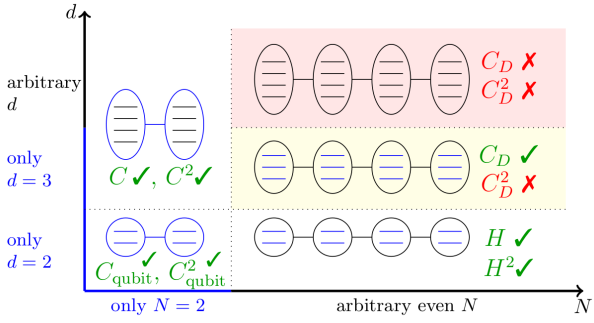}
  \caption{The local unitary invariant $C_D$, cf.~Eq.~\eqref{eq:cdquadrat},
           termed distributed concurrence, naturally generalizes
           the well-known cases of bipartite concurrence 
           $d\geqq 2$, $N=2$ \cite{Rungta2001,Albeverio2001}
           and $H$ invariant for $N$-qubit system~\cite{Wong2001}.
           While in those cases both $C_D$ and $C_D^2$ are monotones,
           the general case $d>2$, $N>2$ is different: $C_D$ (but {\em not}
           $C_D^2$) is an entanglement monotone only if all local dimensions
           $d\leqq 3$. If there is one subsystem of dimension
           at least four, neither $C_D$ nor $C_D^2$ are entanglement
           monotones.
    }
  \label{fig:monotones}
\end{figure}
%%%%%%%%%%%%%%%%%%%%%%%%%%%%%%%%%%%%%%%%%%%%%%%%%%%%%%%%%%%%%%%%%%%%%
%
For local dimensions $d_j\leqq 3$ and even party number $N$,
$C_D$ is an entanglement monotone, but not $C_D^2$. For local dimensions
$d_j\geqq 4$ neither $C_D$ nor $C_D^2$ is an entanglement monotone
for any even $N>2$.

%%%%%%%%%%%%%%%%%%%%%%%%%%%%%%%%%%%%%%%%%%%%
%%%%%%%%%%%%%%%%%%%%%%%%%%%%%%%%%%%%%%%%%%%%
%

%%%%%%%%%%%%%%%%%%%%%%%%%%%%%%%%%%%%%%%%%%%%%%%%%%%%%%%%%%%%%%%%%%%%%%
\section{Distribution of correlations and monogamy of entanglement}
\label{sec:monogamy}
%%%%%%%%%%%%%%%%%%%%%%%%%%%%%%%%%%%%%%%%%%%%%%%%%%%%%%%%%%%%%%%%%%%%%%

In Ref.~\cite{ES2015}, the monogamy relation Eq.~\eqref{eq:monog2}
for $N$ qubits was derived. The strategy there was to write down
a local SL invariant for an $N$-qubit state $\rho_N$
via combining $\rho_N$ and $\tilde \rho_N$, and, by expanding that
expression in terms of Bloch components to obtain an identity which
connects the invariant (for the global state) with combinations of
local quantities (linear entropies). Now we ask whether one can
obtain a new result by generalizing Eq.~\eqref{eq:monog2} in the frame
of the present formalism. The answer is affirmative, as we will show.

We consider a state $\rho$ of a finite-dimensional multipartite 
quantum system 
$\rho \in \mathcal{B}(\mathcal{H}_1\otimes\ldots\otimes\mathcal{H}_N)$.
Clearly, $\Tr\left(\rho\tilde \rho\right)$ is a local unitary invariant.
We expand this expression as a sum over all bipartite splits $A|\bar{A}$
by applying Eq.~\eqref{eq:rhotilde-reduced}
\begin{align}
   \Tr\left(\rho\rhoti\right)\ & =\ 1+\sum_{|A|>0}   (-1)^{|A|}  
                           \Tr\Big[ \rho \Tr_{\bar A} \rho\otimes\id_{\bar A}
                              \Big]
\nonumber\\
                               & =\ 1+\sum_{|A|>0}   (-1)^{|A|} 
                           \Tr_A\Big[ \Tr_{\bar A} \rho \Big]^2
\nonumber\\
                               & =\ \frac{1}{2}
                               \sum_{|A|>0} (-1)^{|A|+1}2\Big(1-\Tr\rho_A^2\Big)
\nonumber
\end{align}
By using the definition of the linear entropy on the partition $A$,
$\tau_A=2(1-\Tr\rho_A^2)$ we obtain
\begin{align}
 0\ \leqq\  2 \Tr\left(\rho\rhoti\right)\  =
                               \sum_{|A|>0} (-1)^{|A|+1}\ \tau_A
\ \ .
\label{eq:monomixed}
\end{align}
This is an equality for the distribution of correlations and holds 
both for pure and for mixed states of any 
finite-dimensional multipartite system. 
The general validity of Eq.~\eqref{eq:monomixed} is quite remarkable: 
In Ref.~\cite{ES2015} it was derived by using a qubit-specific formalism,
therefore its full generality could not be expected.

It is in order here to recall the quantum marginal 
     problem~\cite{Klyachko2006}, which poses the question
     whether or not a given set of
     marginal states is compatible with a (possibly unique) global
     state. The equality~\eqref{eq:monomixed} represents a simple necessary
     (but not sufficient) condition for a favorable answer to that
     question.

It is interesting here to consider the special case of 
mixed states $\rho_{ABC}$ of {\em three-parties} A, B, and C.
From Eq.~\eqref{eq:monomixed} and $\Tr(\rho\rhoti)\geqq 0$ we find
\begin{align}
\tau_A+\tau_B+\tau_C+\tau_{ABC}\ \geqq\ \tau_{AB}+\tau_{AC}+\tau_{BC}
\ \ .
\label{eq:stronglin}
\end{align}
This reminds of the {\em strong subadditivity
inequality} for the von-Neumann entropy $S$~\cite{Lieb1973}, which reads
$S_{ABC}+S_B \leqq S_{AB}+S_{BC}$. However, inequality~\eqref{eq:stronglin}
is symmetrized between the parties and has the opposite direction.

We emphasize that the $\tau_A$ in Eq.~\eqref{eq:monomixed}
are simply the linear entropies of the reduced states $\rho_A$, so they
are {\em not} entanglement measures. However, if we consider {\em pure}
states $\ket{\psi}\!\bra{\psi}\equiv\Pi_{\psi}$, we note
$\tau_A(\psi)=0$ for $|A|=N$, and the equality reads
\begin{align}
  0\ \leqq\ 2 C_D^2(\psi)\  & = \ \sum_{|A|>0} (-1)^{|A|+1}\ \tau_A(\psi)
\nonumber\\
                  & = \ \sum_{N>|A|>0} (-1)^{|A|+1}\ C_{A|\bar A}^2(\psi)\ .
\label{eq:monopure}
\end{align}
Here we denote by $C_{A|\bar A}(\psi)$ the bipartite concurrence 
on the split $A|\bar A$ according to Eq.~\eqref{eq:inv2conc2}.
We see that Eq.~\eqref{eq:monopure} represents a true monogamy relation,
because all the terms on the right-hand side are entanglement measures. 
If the local dimensions
are not larger than three, also $C_D$ is an entanglement measure, so
that we even have a monogamy equality.
Thus, Eq.~\eqref{eq:monopure} is a constraint for the distribution
of entanglement across all bipartitions of an arbitrary multipartite
pure state of finite dimensionality. 
For odd $N$ the equality is trivial, because the purities on 
complementary partitions $A$ and $\bar A$
enter it with a different sign, confirming
that $C_D(\psi)=0$ in this case. 
     We note that aspects of the results discussed here and  summarized in 
     Eqs.~\eqref{eq:monomixed}, \eqref{eq:monopure} were obtained in 
     Ref.~\cite{Cai2007}.

It is an interesting feature of Eqs.~\eqref{eq:monopure},~\eqref{eq:stronglin}
that they describe all local parties in a symmetric manner, whereas
the Coffman-Kundu-Wootters identity Eq.~\eqref{eq:CKW} or the 
strong subadditivity inequality single
out one party. Our way of deriving the general monogamy 
relation~\eqref{eq:monopure} seems to suggest that the
symmetric form has a peculiar role, just because the 
tilde state $\rhoti$ is symmetric in the local parties as well.
%In fact, if the definition of $\rhoti$ in Eq.~\eqref{eq:rhotildeprod}
%did not contain all local parties, the resulting operator would not
%necessarily be positive.
%%%%%%%%%%%%%%%%%%%%%%%%%%%%%%%%%%%%%%%%%%%%%%%%%%%%%%%%%%%%%%%%%%%%%
%% FIGURE 3
%%%%%%%%%%%%%%%%%%%%%%%%%%%%%%%%%%%%%%%%%%%%%%%%%%%%%%%%%%%%%%%%%%%%%
\begin{figure}[thb]
  \centering
  \includegraphics[width=.90\linewidth]{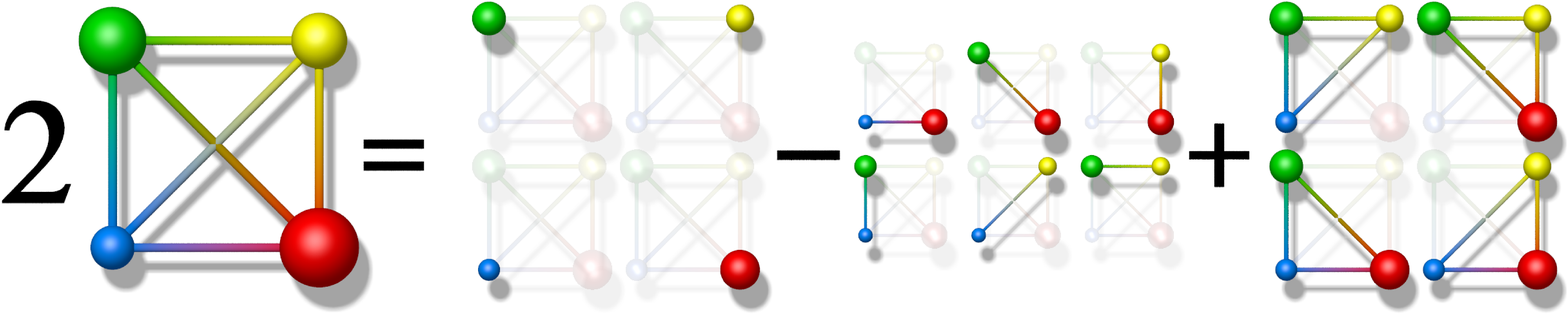}
  \includegraphics[width=1.\linewidth]{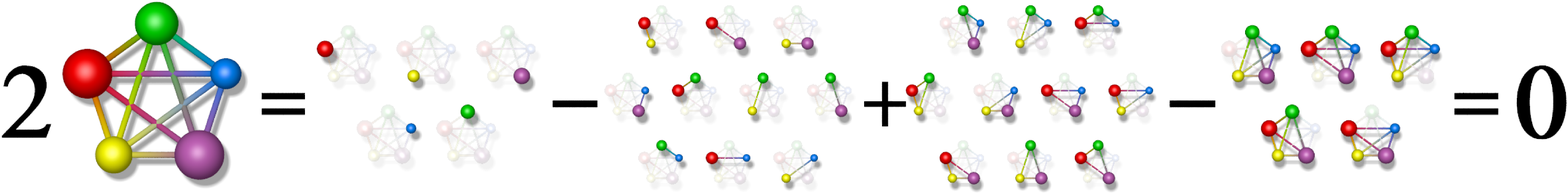}
  \caption{Illustration of the $N$-qudit monogamy relation 
           Eq.~\eqref{eq:monopure}. 
           The left-hand side symbolizes $C_D^2$ while, on 
           the right-hand side, there are the representations
           of the linear entropies of all possible bipartite splits
           in the set of $N$ qudits.\protect\newline 
           Top panel: Four qudits.
           Complementary partitions
           $A$ and $\bar A$ enter with the same sign.\protect\newline
           Bottom panel: Five qudits. For odd $N$, the monogamy
           relation is trivial because the linear entropy
           of a partition $A$ equals that of its complementary
           partition $\bar A$, but contributes with opposite sign.
    }
  \label{fig:monogamy}
\end{figure}
%%%%%%%%%%%%%%%%%%%%%%%%%%%%%%%%%%%%%%%%%%%%%%%%%%%%%%%%%%%%%%%%%%%%%
%

\section{Other interesting aspects}

\subsection{Conservation laws for correlations }

In the previous section we have emphasized that in Eq.~\eqref{eq:monomixed}
all terms are local unitary invariants. The left-hand side is locally
invariant on each single party, whereas the terms on the right-hand side
are locally invariant on their respective bipartition. We explain now
that Eq.~\eqref{eq:monomixed} gives rise to nontrivial unitary invariants 
on an entire subset of the parties.

To this end, let us consider an example of a four-party state $\rho_{1234}$
where each index denotes the respective party. The reduced state, e.g.,
of parties $\{13\}$ is $\rho_{13}=\Tr_{24}\rho_{1234}$. Correspondingly,
the linear entropy of parties $\{13\}$ is 
$\tau_{13}=2(1-\Tr_{13}\rho_{13}^2)$. Then we have
\begin{align}
   \Tr\rho_{1234}\tilde{\rho}_{1234}
   \ =\ & \tau_1+\tau_2+\tau_3+\tau_4-\tau_{12}-\ldots -
\nonumber\\
    &\!\!\!\!\! -\tau_{34}+\tau_{123}+\tau_{124}+\tau_{134}+\tau_{234}-
\nonumber\\
    &\!\!\!\!\! -\tau_{1234} \ .
\label{eq:ex4}
\end{align}
Assume now that the subsystem of parties $\{13\}$ is subject to
unitary evolution 
$U_{13}(t)=\id_{24}\otimes\exp\left(-\rmi \mathsf{H}_{13}t\right)$
generated by a (time-independent) 
Hamiltonian $\mathsf{H}_{13}$. The full state $\rho_{1234}$
be mixed, e.g., the thermal state at some temperature. Let us collect 
all the terms in Eq.~\eqref{eq:ex4} in which both of the parties 1 and 3,
or none of them, are traced out:
\begin{align}
  -\tau_2 & -\tau_4+\tau_{13}+ \tau_{24}-\tau_{123}-\tau_{134} +\tau_{1234}\ =\  
\nonumber\\
  =\  &-\Tr\rho_{1234}\tilde{\rho}_{1234}+\tau_1+\tau_3- \tau_{12}-
      \tau_{14}-
\nonumber\\
    & -\tau_{23}-\tau_{34}+\tau_{124}+\tau_{234}
\label{eq:ex42}
\end{align}
The left-hand side of Eq.~\eqref{eq:ex42} is invariant under
the evolution $U_{13}(t)$ while each of the terms on the right will
vary in a nontrivial way, as the time evolution generates or 
degrades correlations of the corresponding parties with the other
subsystems. This means Eq.~\eqref{eq:ex42}
represents a {\em conservation law for correlations}.
Analogous conservation laws can be found for Hamiltonian subsystem dynamics
of any finite-dimensional multipartite state.

\subsection{Relation between distributed concurrence and topology}
\label{sec:other-Top}

A striking property of Eqs.~\eqref{eq:monomixed} and~\eqref{eq:monopure}
is their apparent connection with topology. Just the form of those 
equations hints at the so-called Euler characteristic~\cite{Crossley},
a number characterizing the structure of a topological space. Yet in the
present case the nature of the topological entity, described
in particular by Eq.~\eqref{eq:monopure} is not really clear. Another 
problem is that Eq.~\eqref{eq:monopure} in general is not a relation
between integers, which makes its topological interpretation questionable.

However, there are more indications in the properties of $C_D$ that there 
indeed exists a relation to topology. First, it is known that a closed
topological space of odd dimension has Euler characteristic 0. This 
coincides with the property $C_D\equiv 0$ for systems with odd number
of parties $N$. Further, the Euler characteristic 
of a product manifold equals the product of the Euler characteristics of
the factors. Again, it is easy to see from the explicit definition of
$C_D$, Eq.~\eqref{eq:CDy}, that the distributed concurrence has the
same property. Finally, the Euler characteristic bears a close relation
with the inclusion-exclusion principle. The derivation 
of Eq.~\eqref{eq:monopure} for qubits in Ref.~\cite{ES2015}
via the Bloch representation is an explicit application of this principle.

We can only speculate what the topological nature of Eq.~\eqref{eq:monopure}
is. A tempting interpretation is the following. We regard the $N$ parties
as vertices of a standard $(N-1)$-simplex. 
In a simplex, each vertex is connected
to each other vertex. This appears to correspond to the distribution
of entanglement in a genuinely entangled~\cite{DVC2000} pure state: 
A given party is entangled with any other party in such a pure state. 
In a strict sense there
is no distinction, such as `only pairwise entanglement', as is possible
for mixed states. The boundary of an \mbox{$(N-1)$}-simplex is 
topologically equivalent
to a sphere $S^{N-2}$ and therefore has Euler characteristic $1+(-1)^{N-2}$
which corresponds to the behavior of $C_D$ (vanishing $C_D$ for odd $N$).  
Finally, if the pure state can
be written as a product of exactly two factors (with party numbers 
$k$ and $N-k$), the topology would be that
of two disjoint simplices, one with $k$ and the other with $N-k$
vertices. 

\section{Conclusions}
We have analyzed universal state inversion applied to
multipartite quantum states with finite-dimensional parties.
This map takes pure states to the opposite hypersurface of the
state space and therefore algebraically encodes extremal properties
of quantum states. We %also 
made explicit that this map is antilinear.
In the framework of our formalism we were able to derive a universally
valid equality, Eq.~\eqref{eq:monomixed},  that relates 
a local unitary invariant depending on the global state
to the linear entropies of all possible reduced states. Remarkably,
if we apply this equality to pure states we obtain 
the first known 
monogamy relation for bipartite entanglement in a multipartite state,
Eq.~\eqref{eq:monopure}, which is valid for any 
number of parties and any finite local dimension. 
These findings appear to have a close relation
with the topological properties of multipartite quantum systems, as
we have argued in Sec.~\ref{sec:other-Top}. Moreover, our correlation
equalities give rise to conservation laws for correlations, which 
are valid for any type of unitary subsystem dynamics.

The local unitary invariant mentioned above, the distributed concurrence,
describes the quantitative distribution of bipartite entanglement 
in a multipartite pure state.  It naturally generalizes
the concurrence in the well-known cases, that is, for bipartite systems
of arbitrary dimension
and for many-qubit systems. Thus, it provides both the mathematical 
and conceptual link between these hitherto somewhat unrelated
quantities, see Eqs~\eqref{eq:monog2},\eqref{eq:inv2conc2}. 
Intriguingly,  for systems of even party number
$N\geqq 4$ the distributed concurrence is an entanglement monotone only below 
a `critical' local dimension $d\leqq 3$.  Note, however, that for
its role as a `generator' of the monogamy inequality the monotone property
of the distributed concurrence is not essential.

Our work shows that, by means of algebraic methods, 
it is possible to express the mathematical constraints to quantum states,
in particular their positivity, in terms of physically relevant 
quantities and principles. Importantly, our results establish necessary
relations between the global and the local states which may be considered
as a partial answer in the context of the quantum-marginal problem.
Clearly, our result highlights only one specific aspect of correlations,
and the question is whether also other mathematical constraints 
allow for a similar formulation with direct access to their physical meaning.

%%%%%%%%%%%%%%%%%%%%%%%%%%%%%%%%%%%%%%%%%%%%
%\emph{Acknowledgements}.---
\acknowledgments
%%%%%%%%%%%%%%%%%%%%%%%%%%%%%%%%%%%%%%%%%%%%
This work was funded by the German Research Foundation Project
EL710/2-1 (C.E.), by Basque Government grant IT986-16,
MINECO/FEDER/UE grants FIS2012-36673-C03-01  and  FIS2015-67161-P
and UPV/EHU program UFI 11/55 (J.S.). 
The authors would like to thank Jaroslav Fabian, Felix Huber,  Marcus Huber,
Peter D.\ Jarvis, and Denis Kochan  for stimulating discussions, and
Klaus Richter for supporting this project. F.H.\ pointed out to us
Refs.~\cite{Hall2005,Hall2006,Cai2007}. 
D.K.\ noticed the apparent similarity between
the monogamy relation and the Euler characteristic.
%
%\section{Supplemental Material}
%%\input{supplemental.tex}
%\input{suppj.tex}
%
\appendix
%

%

%
%%%%%%%%%%%%%%%%%%%%%%%%%%%%%%%%%%%%%%%%%%%%%%%%%%%%%%%%%%%%%%%%%%%%%%%
\section{Proof of useful generator formulae}
%%%%%%%%%%%%%%%%%%%%%%%%%%%%%%%%%%%%%%%%%%%%%%%%%%%%%%%%%%%%%%%%%%%%%%%
%
The interesting relations~\eqref{eq:tracefromgenerators}, 
\eqref{eq:transposefromgenerators} are closely linked to
the well-known completeness relation for the set of generators $\{\hd_k\}$
(see, e.g.,~\cite{Georgi1982}). It is instructive to derive all these 
relations in the same framework, i.e., by consequently using the
terminology of quantum information.

An important ingredient is the $\swap$ operator. Consider two
states $\ket{\psi}$, $\ket{\phi}\in\mathcal{H}$. Then we define
\begin{align}
   \swap\cdot\left(\ket{\psi}\otimes\ket{\phi}\right)\ =\ 
               \ket{\phi}\otimes\ket{\psi}
\ \ .
\end{align}
With the computational basis $\{\ket{j}\}$ of $\mathcal{H}$ it is
straightforward to show that $\swap=\sum_{jk}\ket{jk}\!\bra{kj}$.
Let us denote by $\Tr_{[1]}$ the partial trace over the first 
tensor factor in $\mathcal{H}\otimes\mathcal{H}$, and correspondingly
by $\Tr_{[2]}$ the partial trace over the second factor. Then we
find by explicit calculation in the computational basis
that for two operators 
$A$, $B\in \mathcal{B}(\mathcal{H})$
\begin{align}
   \Tr_{[1]}\left(\swap\cdot (A\otimes B\right))\ & =\ A\cdot B\ \ 
\label{eq:swapAB1}
\\
   \Tr_{[2]}\left(\swap\cdot (A\otimes B\right))\ & =\ B\cdot A
\label{eq:swapAB2}
\end{align}
and
\begin{align}
   \Tr_{[1]}\left((A\otimes B\right)\cdot \swap)\ & =\ B\cdot A\ \ 
\label{eq:swapAB3}
\\
   \Tr_{[2]}\left((A\otimes B\right)\cdot \swap)\ & =\ A\cdot B\ \ .
\label{eq:swapAB4}
\end{align}
Important special cases of these relations are obtained by setting, e.g.,
$B=\id$,
\begin{align}
   \Tr_{[1]}\left(\swap\cdot (A\otimes \id\right))\ & =\ A\ \ .
\label{eq:trick1}
%\\
%   \Tr_{[2]}\left(\swap\cdot A\otimes \id\right)\ & =\  A\ \  .
\end{align}
Now consider a set $\{\hd_j\}$, $j=1\ldots(d^2-1)$ of traceless 
Hermitian generators
of SU($d$), $\Tr\left(\hd_j\hd_k\right)=d \delta_{jk}$, $\hd_0\equiv\id$, 
so that we can expand
\begin{align}
          A\ =\ & \frac{1}{d}\sum_{j=0}^{d^2-1} a_j\hd_j\ =\ 
                \frac{1}{d}\sum_{j=0}^{d^2-1} \Tr\left(\hd_j A \right)\hd_j 
\\
   =  \ &\Tr_{[1]}\bigg[ \bigg(\frac{1}{d}\sum_{j=0}^{d^2-1} \hd_j\otimes\hd_j
                 \bigg)\cdot \big( A\otimes\id \big)
                 \bigg]\ \ .
\label{eq:expand}
\end{align}
Because $A$ is an arbitrary operator here, comparison of Eqs.~\eqref{eq:trick1},
\eqref{eq:expand} yields the completeness relation for the generators $\hd_j$
\begin{align}
     \swap\ =\ \frac{1}{d}\sum_{j=0}^{d^2-1} \hd_j\otimes\hd_j\ \ .
\label{eq:swapgen}
\end{align}

With these prelimary considerations we can turn to the proof of 
the operator relations in Sec.~\ref{sec:generators}. Inserting 
Eq.~\eqref{eq:swapgen} into the identity
\[
     \swap\cdot\left(A\otimes\id\right)\cdot\swap\ =\ \id\otimes A
\]
and tracing over the second party gives
\begin{align}
  \Tr\left(A\right)\id\ =\ & \frac{1}{d^2}
                               \sum_{j,k=0}^{d^2-1}\left( \hd_j A \hd_k\right)
                         \Tr\left( \hd_j\hd_k \right)
\nonumber\\
            =\ & \frac{1}{d}\sum_{j=0}^{d^2-1} \hd_j A \hd_j\ \ ,
\end{align}
which proves Eq.~\eqref{eq:tracefromgenerators}.

From the explicit representation of $\swap$ in the computational basis 
it is obvious that it is related to the maximally entangled state
$\ket{\Phi^+}\equiv \frac{1}{\sqrt{d}}\sum_j\ket{jj}$ by virtue 
of partial transposition, e.g., on the first party, $T_{[1]}$, 
\begin{align}
    \swap^{T_{[1]}}\ =\ & \left(\sum_{jk} \ket{jk}\!\bra{kj}\right)^{T_{[1]}}    
\nonumber\\
     =\ & \sum_{jk} \ket{kk}\!\bra{jj}
\nonumber\\
     =\ & d\ \ket{\Phi^+}\!\bra{\Phi^+}\ \ .
\label{eq:bellswap}
\end{align}
We conclude from Eqs.~\eqref{eq:swapgen} and~\eqref{eq:bellswap} that
\begin{align}
     \swap^{T_{[1]}}\ =\ 
%& d\ \ket{\Phi^+}\!\bra{\Phi^+}
%\nonumber\\
%                    \  =\ &
          \frac{1}{d}\sum_{j=0}^{d^2-1} \hd_j^T\otimes\hd_j\ \ .
\label{eq:swapTgen}
\end{align}
In analogy with Eqs.~\eqref{eq:swapAB1}--\eqref{eq:swapAB4}
we obtain now (by explicit calculation in the computational basis)
\begin{align}
   \Tr_{[1]}\left(d \ket{\Phi^+}\!\bra{\Phi^+}
                  \cdot (A\otimes B)\right)\ & =\ A^T\cdot B\ \ 
\label{eq:bellAB1}
\\
   \Tr_{[2]}\left(d \ket{\Phi^+}\!\bra{\Phi^+}
                  \cdot (A\otimes B)\right)\ & =\ B^T\cdot A\ \  
\label{eq:bellAB2}
\\
   \Tr_{[1]}\left((A\otimes B)\cdot
                  d \ket{\Phi^+}\!\bra{\Phi^+}
                  \right)\ & =\ B\cdot A^T\ \ 
\label{eq:bellAB3}
\\
   \Tr_{[2]}\left((A\otimes B)\cdot
                  d \ket{\Phi^+}\!\bra{\Phi^+}
            \right)      
                  \ & =\ A\cdot B^T\  .
\label{eq:bellAB4}
\end{align}
By using Eq.~\eqref{eq:bellAB1} with $B=\id$, the
permutation invariance of $\ket{\Phi^+}$, and Eq.~\eqref{eq:swapTgen} we find
\begin{align}
      A^T\ =\ & \Tr_{[1]}\left(
                  d \ket{\Phi^+}\!\bra{\Phi^+}\cdot (A\otimes \id)\right)
\nonumber\\
           =\ & \Tr_{[1]}\left( \swap\cdot d \ket{\Phi^+}\!\bra{\Phi^+}
                \cdot (A\otimes \id)\right)
\nonumber\\
           =\ & \frac{1}{d}\sum_{j=0}^{d^2-1} 
                \Tr_{[1]}\left(\swap\cdot
                \left[(\hd_j^T A)\otimes \hd_j\right]\right) 
\nonumber
\end{align}
and, finally, with Eq.~\eqref{eq:swapAB1}
\begin{align}
      A^T\ 
           =\ & \frac{1}{d}\sum_{j=0}^{d^2-1} 
                \hd_j^T A\hd_j\ \ ,
\end{align}
which proves Eq.~\eqref{eq:transposefromgenerators}.

%%%%%%%%%%%%%%%%%%%%%%%%%%%%%%%%%%%%%%%%%%%%%%%%%%%%%%%%%%%%%%%%%%%%%%%
\section{Proof of monotone property in the known cases
         }
%%%%%%%%%%%%%%%%%%%%%%%%%%%%%%%%%%%%%%%%%%%%%%%%%%%%%%%%%%%%%%%%%%%%%%%
%
\label{app:monotones}

For the sake of completeness, we present the proof for the 
monotone property of both $C_D$ and $C_D^2$ in the known cases
in the frame of the formalism in Sec.~\ref{sec:monotonie}.

{\em $d=2$, $N$ arbitrary:} Here, there are only two indices $j\in\{0,1\}$ and
only one value of $F_{0110}=F_{1001}$, so that the inequality~\eqref{eq:mon2}
 becomes $\sqrt{2 F_{0110}}\geqq 
\sqrt{2 F_{0110}D_0D_1}+
\sqrt{2 F_{0110}(1-D_0)(1-D_1)} $ from which
we have
\begin{align*}
    1\ \geqq\ \sqrt{D_0D_1}\ +\ \sqrt{(1-D_0)(1-D_1)}
\ \,
\end{align*}
and further
\begin{align*}
    \left(1-\sqrt{D_0D_1}\right)^2\ \geqq\ (1-D_0)(1-D_1)
\end{align*}
which is equivalent to
\begin{align*}
    D_0+D_1\ \geqq\ 2\sqrt{D_0D_1}
\end{align*}
and always fulfilled
because of the relation between arithmetic and geometric mean.

The monotone condition Eq.~\eqref{eq:mon1} for the square
of $C_D$ reads
\begin{align}
   C_D^2(\psi)\ & \geqq\ p_1\ C_D^2\left(\frac{\mathcal{A}_1\psi}
                                        {\sqrt{p_1}}
                             \right)\ +\
   p_2\ C_D^2\left(\frac{\mathcal{A}_2\psi}
                                        {\sqrt{p_2}}
                             \right)
\nonumber\\
              & \geqq\ 
   \frac{C_D^2\left(\mathcal{A}_1\psi \right)}{p_1}\ +\
   \frac{C_D^2\left(\mathcal{A}_2\psi \right)}{p_2}
 \ \ .
\label{eq:monsquare}
\end{align}
In the proof for $C_D^2$ we use the probability 
$p_1=\bra{f_0}f_0\rangle D_0 +\bra{f_1}f_1\rangle D_1$
(recall that $\sum_j\bra{f_j}f_j\rangle=1 $),
and we have to show that
\begin{align*}
    1\ \geqq\ \frac{D_0D_1}{p_1}\ +\ \frac{(1-D_0)(1-D_1)}{1-p_1}
\ \ .
\end{align*}
This inequality is equivalent to 
\begin{align*}
 0\ \geqq\ (p_1-D_0)(p_1-D_1)
\end{align*}
which is always fulfilled as is easily checked with the explicit
formula for $p_1$.

{\em $N=2$, $d$ arbitrary:} In this case we have 
\begin{align*}
   F_{jkkj}\ =\ & \bra{f_j}\left(\widetilde{\ket{f_k}\!\bra{ f_k }}
                              \right)
                     \ket{f_j}
\\ 
             =\ & \bra{f_j}\bigg(\Tr\big(\ket{f_k}\!\bra{ f_k }\big)\id
                        -\ket{f_k}\!\bra{ f_k }
                              \bigg)
                     \ket{f_j}
\\ 
             =\ & \bra{f_j}f_j\rangle \bra{f_k}f_k\rangle - 
                      \left| \bra{f_j}f_k\rangle\right|^2
\ \ ,
\end{align*}
because the subnormalized vectors $\{\ket{f_j}\}$ belong to a single party.
and inequality~\eqref{eq:mon2} takes the form
\begin{widetext}
\begin{align*}
\sqrt{1-\sum_{jk} \left| \bra{f_j}f_k\rangle\right|^2} \  \geqq\
         \sqrt{p_1^2-\sum_{jk} \left| \bra{f_j}f_k\rangle\right|^2 
                      D_jD_k} \ +
\\
 +\  \sqrt{(1-p_1)^2-\sum_{jk} \left| \bra{f_j}f_k\rangle\right|^2 
                      (1-D_j)(1-D_k)} &
\ \ .
\end{align*}
\end{widetext}
The right-hand side (rhs) of this relation can be upper-bounded by
using the concavity of the square root. Subsequent squaring yields
\begin{widetext}
\begin{align*}
   \left(\mathrm{rhs}\right)^2\ & \leqq\  1-\frac{
               \sum_{jk}\left| \bra{f_j}f_k\rangle\right|^2 D_jD_k}{p_1} 
%\\   & \ \ \ \ \ \
     -\frac{\sum_{jk}\left| \bra{f_j}f_k\rangle\right|^2 (1-D_j)(1-D_k)}{1-p_1}
\\& %\!\!\!\!  \!\!\!\!\!    \!\!\!\!\!
      \leqq\ 1-\sum_{jk} \left| \bra{f_j}f_k\rangle\right|^2 
               \frac{p_1(1-p_1)+(p_1-D_j)(p_1-D_k)}{p_1(1-p_1)}\ \ .
\end{align*}
\end{widetext}
Hence, in order to prove the monotone property it remains to show that
\begin{align*}
    0\ \leqq\           \sum_{jk}\left| \bra{f_j}f_k\rangle\right|^2 
                     (p_1-D_j)(p_1-D_k)\ \ .
\end{align*}
This can be seen by noting that 
\begin{align*}
               \sum_{jk}\left| \bra{f_j}f_k\rangle\right|^2 
                     (p_1-D_j)(p_1-D_k)\ = \ &\Tr\left(W^{\dagger}W\right)
\\              \geqq \ & 0
\end{align*}
with the Hermitian operator
\begin{align*}
                W\ =\ \sum_{j} \ket{f_j}\!\bra{f_j}(p_1-D_j)\ \ ,
\end{align*}
which concludes the proof.
We note that application of the concavity of the square root above
implies that also $C_D^2$ is a monotone in this case.

%
%%%%%%%%%%%%%%%%%%%%%%%%%%%%%%%%%%%%%%%%%%%%%%%%%%%%%%%%%%%%%%%%%%%%%%%
\section{Examples for states violating the monotone property
         }
%%%%%%%%%%%%%%%%%%%%%%%%%%%%%%%%%%%%%%%%%%%%%%%%%%%%%%%%%%%%%%%%%%%%%%%
%
\label{app:counterex}

As we have mentioned in Sec.~\ref{sec:monotonie}, it is possible
to find examples violating the monotone conditions, 
Eqs.~\eqref{eq:mon2}, \eqref{eq:mon3},
which implies that $C_D$ cannot be an entanglement monotone for the 
local dimension in question, and all higher dimensions as well.

A numerical strategy for finding such counterexamples for $d=4$
is the following.
We can restrict our attention to channels which are diagonal in 
the Schmidt basis of the first party. Then, the decomposition
$\ket{\psi}=\sum_j \ket{e_j f_j}$ from Sec.~\ref{sec:monotonie} becomes
the Schmidt decomposition $\ket{\psi}=\sum_j\sqrt{\lambda_j}\ket{e_j\bar{f}_j}$
with orthonormal $\{\ket{\bar{f}_j}\}$ and $\sum_j\lambda_j=1$. 
A strategy to look for parameters to violate Eq.~\eqref{eq:mon3} is to 
construct a matrix $F_{kjjk}\equiv\lambda_j\lambda_k \bar{F}_{kjjk}$ 
(where 
$\bar{F}_{kjjk}\equiv \bra{\bar{f}_k}\left(\widetilde{\ket{\bar{f}_j}\!\bra{\bar{f}_j}}\right)\ket{\bar{f}_k}$)
with sufficiently extremal entries,
that is, with values close to 0 and others of order 1. 

A matrix $\bar{F}_{kjjk}$ with such extremal entries can be found
as follows. First, we generate a state $\psi_1$ with random entries and 
apply the state inverter to the projector $\Pi_{\psi_1}$. 
Since we consider an odd number of parties, the state $\psi_1$ is an 
eigenstate of $\tilde{\Pi}_{\psi_1}$ with eigenvalue 0. In general,
the other eigenvalues of $\tilde{\Pi}_{\psi_1}$ are distributed between
0 and numbers of order 1, so that we can choose three of those other
eigenstates of $\tilde{\Pi}_{\psi_1}$  with suffienctly different
eigenvalues. Once we have computed the matrix $\bar{F}_{kjjk}$
we can perform a random search for Schmidt coefficients $\{\lambda_j\}$
and diagonal entries of the channel $\{D_j\}$ and readily find
numerical examples that violate
the monotone condition for $C_D$ or $C_D^2$. The same procedure
can be applied for $d=3$ in order to find a counterexample for the
monotone condition of $C_D^2$.

To conclude this part, we give an explicit analytical counterexample
for a four-party state (with local dimensions 4, 2, 2, 2)
that violates the monotone condition both
for $C_D$ and $C_D^2$. It reads
\begin{align*}
\ket{\psi_{4222}}\ =\ & \frac{1}{\sqrt{8}}
                 \left(\ket{0000}+\ket{0011}+\ket{1100} +
                 \right.
\\
                      &\ \ \ \   +\ket{1111}+\ket{2000} - \ket{2011} -
\\               & \left.  \ \ 
\ \                       -\ket{3100}+\ket{3111}
                 \right)
 \ \ .
\end{align*}
The local channel acting on the first party is given by the projectors
\begin{align*}
    \mathcal{A}_1\ =\ & \diag{(1,1,0,0)}
\\
    \mathcal{A}_2\ =\ & \diag{(0,0,1,1)}
\ \ .
\end{align*}

%%%%%%%%%%%%%%%%%%%%%%%%%%%%%%%%%%%%%%%%%%%%%%%%%%%%%%%%%%%%%%%%%%%%%%%

%

\end{document}